\documentclass[twocolumn,aps,prl,superscriptaddress,amsfonts,amssymb,amsmath,showpacs]{revtex4}
\usepackage{graphicx}
\begin{document}
\title{\bf The minimum model for the iron-based superconductors}

\author{Jian Li}
\affiliation{Beijing National Laboratory for Condensed Matter
Physics, Institute of Physics, Chinese Academy of Sciences,
Beijing 100190, PR China}
\author{Yupeng Wang$^*$}
\affiliation{Beijing National Laboratory for Condensed Matter
Physics, Institute of Physics, Chinese Academy of Sciences,
Beijing 100190, PR China}

\begin{abstract}
A single band  $t$-$U$-$J_1$-$J_2$ model is proposed as the
minimum model to describe the superconductivity of the newly
discovered iron-based superconductors $R(O_{1-x}F_x)FeAs$ and
$RO_{1-x}FeAs$ ($R=La, Ce, Sm, Pr, Nd,Gd$). With the mean-field
approach, it is found that the pairing occurs in the $d$-wave
channel. In the likely parameter region of the real materials, by
lowering temperature, the system enters firstly the $d_{xy}$
superconducting phase with $D_{4h}$-symmetry and then enters the
time-reversal-symmetry-broken $d_{xy}+id_{x^2-y^2}$
superconducting phase with $C_{4h}$-symmetry.
\end{abstract}


\maketitle

\par
It is well known that the most exciting finding after cuprates in
the family of superconductors is the discovery of
$R(O_{1-x}F_x)FeAs$ ($R=La, Ce, Sm, Pr, Nd, Gd$). Soon after the
announcement of the 26K superconductivity in $F$-doped $LaOFeAs$ by
Kamihara \emph{et al} \cite{hosono}, several groups discovered a
number of compounds with superconducting transition temperatures up
to 52K\cite{ nanlin1,xianhui, zhao1,haihu}. Interestingly, even
without $F$, those compounds may show superconductivity by
introducing some oxygen vacancies\cite{zhao2}. It is argued that
this family of superconductors may share some common features with
the cuprate superconductors and a number of
models\cite{hirsh,zidan,zhong,lee,shoucheng,si} have been proposed
to account for the mechanism of its superconductivity. Band
structure calculations\cite{singh,zhong1,kotliar,zhongyi} show that
six $d$ bands of totally ten $d$ bands cross the Fermi surface but
another group claimed that the parent compound could be a bad
metal\cite{japan} with quite low density of states at the Fermi
surface. A sound experiment combined with computation show that in
the parent compound there is a phase transition from normal metal to
stripe-type spin density wave (SDW) around 150K\cite{nanlin2}. This
SDW state was subsequently demonstrated by the neutron scattering
experiments\cite{pengcheng,manzos}. A theoretical analysis\cite{si}
and \emph{ab initial} computation\cite{zhongyi} indicate that the
stripe SDW phase is induced by the frustrated spin-exchange
interactions and the possible pairing symmetry is $d_{xy}$. These
observations somehow hint that the antiferromagnetic fluctuation may
play an important role for the superconductivity.

Although the band calculations indicate that six bands (hybridized
2$d_{xy}, 2d_{yz}, 2d_{zx}$) cross the Fermi surface, the carriers
very likely come from those bands lying in the a-b plane, i.e.,
those composed of $d_{xy}$- and $d_{x^2-y^2}$-orbits. However, the
numerical result\cite{zhong3} strongly suggests that the
$d_{x^2-y^2}$ band is far above the Fermi surface and $d_{xy}$-orbit
indeed contribute a major weight to the Fermi pocket. The quasi
one-dimensional $d_{yz}$ and $d_{zx}$ bands, though may extend
slightly in the iron plane through hybridization with $p$-orbits of
$As$, are unlikely to dominate the superconductivity. The above
arguments directly lead to the hypothesis that a single band model
may qualitatively describe the mechanism of the superconductivity in
this family of compounds. Therefore, in this Letter, we propose the
following Hamiltonian as the minimum model to account for the
superconductivity in electron doped $ROFeAs$:
\begin{eqnarray}
H=-t\sum_{\sigma,<i,j>}
C^\dagger_{i,\sigma}C_{j,\sigma}+U\sum_{j}n_{j,\uparrow}n_{j,\downarrow}\nonumber\\
+J_1\sum_{<i,j>}\vec{S}_i\cdot\vec{S}_j+J_2\sum_{<<i,j>>}\vec{S}_i\cdot\vec{S}_j,
\end{eqnarray}
where $\sigma$ indicates the spin indices; $C_{j,\sigma}^\dagger$
($C_{j,\sigma}$) is the creation (annihilation) operator of
electrons; $n_{j,\sigma}$ is the particle number operator and
$<i,j>$ and $<<i,j>>$ indicate nearest neighbor and next nearest
neighbor, respectively; $U$ describes the Hubbard repulsion and
$J_{1,2}>0$ represent the exchange constants. The recent ab initial
calculation\cite{zhongyi} suggested that $J_2$ is considerably
larger than $J_1$ due to the superexchange processes through $As$
atoms.

We note that the Hubbard repulsion $U$ is relatively small in the
$FeAs$ compounds compared to those in the cuprates. The
correlation effect is not very strong  and the $FeAs$
superconductors may share some common feature to the so-called
gossamer superconductivity\cite{laughlin,fuchun}. $U$ may have two
major effects on the superconductivity: One is to expel the
$s$-wave pairing\cite{liu} and the other is to renormalize the
band width and the coupling constants $J_{1,2}$\cite{fuchun}. In
such a sense, instead of dealing with model (1) we study the
following Hamiltonian but keep in mind that $t$ and $J_{1,2}$ are
the renormalized constants:
\begin{eqnarray}
H_{eff}&=&-t\sum_{\sigma,<i,j>}
C^\dagger_{i,\sigma}C_{j,\sigma}-J_1\sum_{<i,j>}b_{i,j}^\dagger
b_{i,j}\\ \nonumber &-&J_2\sum_{<<i,j>>}b_{i,j}^\dagger b_{i,j},
\end{eqnarray}
where
\begin{eqnarray}
b_{i,j}=[C_{i,\uparrow}C_{j,\downarrow}-C_{i,\downarrow}C_{j,\uparrow}]
\end{eqnarray}
describes the electron pair operator.

\emph{Mean-field Hamiltonian}.  Because of positive $J_{1,2}$ in
the system, the electrons favor to form spin-singlet pairs. By
introducing the $d$-wave pairing order parameters
$d_1=<b_{i,i+{\hat x}}>=-<b_{i,i+{\hat y}}>$, $d_2=<b_{i,i+{\hat
x}+{\hat y}}>=-<b_{i,i-{\hat x}+{\hat y}}>$, we obtain the the
following mean-field Hamiltonian:
\begin{eqnarray}
H_{\emph{mf}}&=&\sum_kE(k)(\alpha_{k}^\dag\alpha_{k}+\alpha_{-k}^\dag\alpha_{-k})\nonumber\\
&+&\sum_{k}(\xi_{k}-E)+2J_1d_{1}^{2}N+2J_2d_{2}^{2}N,
\end{eqnarray}
with $N$ the number of sites and
\begin{eqnarray}
E(k)&=&\sqrt{\xi_{k}^{2}+|\Delta_{1k}+\Delta_{2k}|^{2}},\nonumber\\
\xi_{k}&=&-2t(cosk_x+cosk_y)-\mu,\nonumber\\
\Delta_{1k}&=&2J_1d_1\gamma_{k},\\
\Delta_{2k}&=&\Delta_{2}=2J_2d_2\eta_{k},\nonumber
\end{eqnarray}
 where $\gamma_{k}=cosk_{{x}}-cosk_{{y}}$ ,
 $\eta_{k}=2\sin k_x\sin k_y$ and $\mu$ denotes the chemical
 potential. The Gibbs free energy at a finite temperature $T$ then
 reads
\begin{eqnarray}
G(T)&=&-2K_BT\sum_kln[2cosh(\frac{E(k)}{2K_BT})]\nonumber\\
&+&\sum_k\xi_k+2J_1d_{1}^{2}N+2J_2d_{2}^{2}N.
\end{eqnarray}
For a given density of electron number $n$, the order parameters
can be determined self-consistently by the following equations:
\begin{eqnarray}
n&=&1-\sum_{k}\frac{\xi_{k}}{E(k)}tanh(\frac{E(k)}{2K_BT}),\nonumber\\
1&=&\frac{J_1}{N}\sum_{k}\frac{\gamma_{k}^{2}}{E(k)}tanh(\frac{E(k)}{2K_BT}),\\
1&=&\frac{J_2}{N}\sum_{k}\frac{\eta_{k}^{2}}{E(k)}tanh(\frac{E(k)}{2K_BT}).\nonumber
\end{eqnarray}
\begin{figure}[htb]
\begin{center}
\includegraphics[width=3.5in]
%[angle=0,height=8cm,width=5.2cm,scale=1.0,bb=75 403 275 592]
%\includegraphics[angle=0,height=5cm,width=5.2cm,scale=1.0,bb=75 403 275 592]
{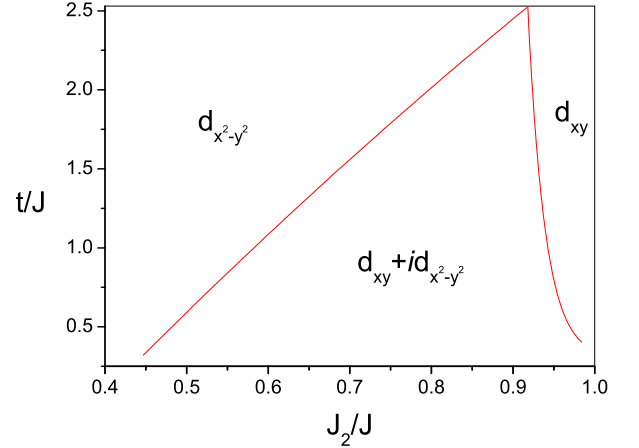} \caption{(color online) The mean-field phase diagram of
the ground state for $n=1.1$. The red lines indicate the phase
boundaries of $d_{x^2-y^2}$, $d_{x^{2}-y^{2}}+\emph{i}d_{xy}$ and
$d_{xy}$ pairing phase.}. \label{fig2}
\end{center}
\end{figure}
\begin{figure}[htb]
\begin{center}
\includegraphics[width=3.5in]
%[angle=0,height=8cm,width=5.2cm,scale=1.0,bb=75 403 275 592]
%\includegraphics[angle=0,height=5cm,width=5.2cm,scale=1.0,bb=75 403 275 592]
{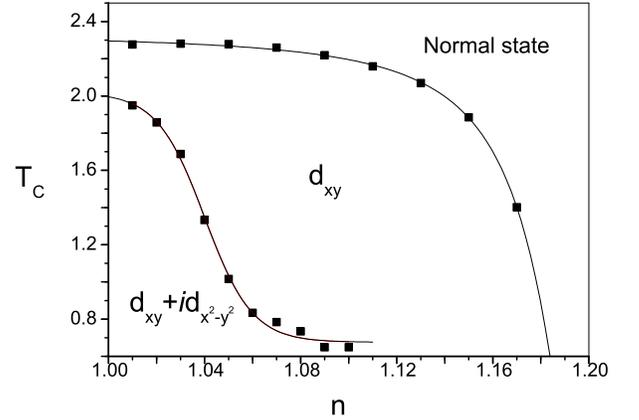} \caption{(color online) Finite temperature phase
diagram for $t=2J$ and $J_2=0.9J$.} \label{fig2}
\end{center}
\end{figure}
\emph{Ground State.} In the ground state, the above
self-consistent equations are reduced to:
\begin{eqnarray}
n&=&1-\sum_{k}\frac{\xi_{k}}{E(k)},\nonumber\\
1&=&\frac{J_1}{N}\sum_{k}\frac{\gamma_{k}^{2}}{E(k)},\\
1&=&\frac{J_2}{N}\sum_{k}\frac{\eta_{k}^{2}}{E(k)}.\nonumber
\end{eqnarray}
Obviously, there are several sorts of solutions for the gap
functions. If $d_2=0$, we get the conventional $d_{x^2-y^2}$ phase
in the ground state; If $d_1=0$, we have the $d_{xy}$ phase which
is isomorphic to the $d_{x^2-y^2}$ phase by rotating the $k$ space
with an angle of $\pi/4$; For both $d_{1,2}\neq 0$, there is an
extra free parameter $\theta$ ($\exp(i\theta)=J_1|J_2|/|J_1|J_2$)
which can not be determined by the self-consistent equations but
can be fixed by the lowest energy. Our numerical solution shows
that only three superconducting phases are possible with the
variation of $t$ and $J_{1,2}$, i.e., the $d_{x^2-y^2}$ phase for
larger $J_1/J_2$, the $d_{xy}$ phase for quite smaller $J_1/J_2$
and the $d_{xy}+id_{x^2-y^2}$ phase with $\theta=\pi/2$ in the
intermediate parameter range. The third phase is quite interesting
because in it the time-reversal symmetry is broken. Its point
group symmetry is also reduced to $C_{4h}$ from $D_{4h}$ of the
$d_{xy}$ state. Of course there are also other kinds of mixed
solutions such as $d_{xy}+d_{x^2-y^2}$, but the corresponding
energy is always higher than that of $d_{xy}+id_{x^2-y^2}$. For
large enough $t$, $d_{1,2}=0$ and the system is in the normal
metallic phase. The mean-field phase diagram of the ground state
for $n=1.1$ and fixed $J=\sqrt{J_{1}^{2}+J_{1}^{2}}$ is depicted
in Fig.1. In the real materials, the band width could be strongly
reduced by the Hubbard repulsion and the value of $J_2$ is around
$2J_1$ or even larger as suggested by the ab initial
calculation\cite{zhongyi}. In this sense, the superconducting
ground state must be either the $d_{xy}+id_{x^2-y^2}$  or the
$d_{xy}$ paired state.

\emph{Finite Temperature Phase Diagram}. For given $t$ and
$J_{1,2}$, the thermodynamic phase diagram can be derived from the
self-consistent equations by taking the order parameters tending to
zero. Fig.2 shows the $T-n$\cite{u} phase diagram for $t=2J$ and
$J_2=0.9J$(Here $n$ does not mean the true carrier density in the
present mean-field approach because we ignore $U$. In the real
materials $T_c$ around $n=1$ could be depressed heavily by $U$). It
is found that there are two superconducting phases. One is the
$d_{xy}$ phase which breaks the $U(1)$ gauge symmetry and the other
is the $d_{xy}+id_{x^2-y^2}$ phase which breaks both the $U(1)$
gauge symmetry and the time-reversal symmetry. The $d_{xy}$ state
has a nodal gap function which allows gapless excitations and
therefore gains entropy at finite temperatures, while the
$d_{xy}+id_{x^2-y^2}$ state is fully gapped with lower energy. The
existence of two superconducting phases is due to the competing
effect between energy and entropy. It is emphasized the second
thermodynamic phase transition could be obtained from the specific
heat measurement.

\emph{Density of States}. One of the important quantities is the
low temperature density of states, which can be detected directly
by the local probe tunnelling experiments. In our
$d_{xy}+id_{x^2-y^2}$ case, it reads :
\begin{eqnarray}
\frac{\rho(\omega)}{N_F}=\int\frac{d\varphi}{2\pi}Re\frac{\omega}{\sqrt{\omega^2-|\Delta_{1}|^2cos^2(2\varphi)-|\Delta_{2}|^2sin^2(2\varphi)}},
\end{eqnarray}
\begin{figure}[htb]
\begin{center}
\includegraphics[width=3.5in]
%[angle=0,height=8cm,width=5.2cm,scale=1.0,bb=75 403 275 592]
%\includegraphics[angle=0,height=5cm,width=5.2cm,scale=1.0,bb=75 403 275 592]
{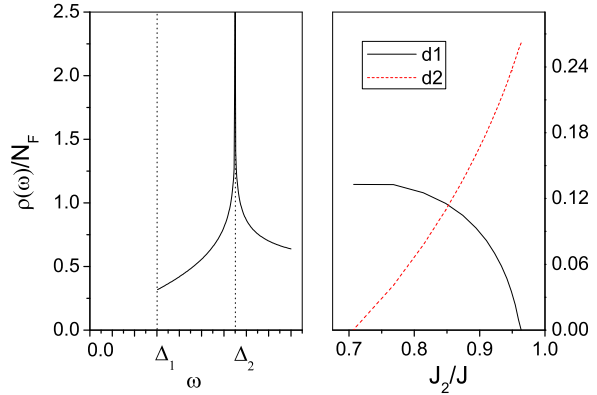} \caption{(color online) The left figure is the density
of states of the $d_{x^2-y^2}+\emph{i}d_{xy}$ ground state. The
right one is the order parameters $|d_{1,2}|$ with the variation
of $J_2/J$ } \label{fig2}
\end{center}
\end{figure}
where $N_F$ is the density of states of the normal phase at the
Fermi surface; $\Delta_{1,2}=2J_{1,2}d_{1,2}$. The numerical
result of $\rho(\omega)$ is depicted in Fig.3. Unlike that of the
$d_{x^2-y^2}$ superconductors, there is no node in this
time-reversal-symmetry-broken superconductor. However, the gap
function is strongly anisotropic with a small minimum gap of
$|\Delta_1|$ in the spectrum.

In conclusion, a minimum model to account for the mechanism of the
iron-based superconductors is proposed. With the mean-field
approach, it is found that the most likely superconducting ground
state is the $d_{xy}$ or $d_{xy}+id_{x^2-y^2}$ type. At finite
temperature, the $d_{xy}$ superconducting state appears first and a
second phase transition into the $d_{xy}+id_{x^2-y^2}$
superconducting state occurs in some doping region by lowering the
temperature. We emphasize that though the real systems are
multi-band superconductors, we believe that our single band model
captures the central physics for the mechanism of the
superconductivity. The multi-band structure may only induce multi
gaps or renormalizaion of the critical temperature without affecting
the mechanism.

We acknowledge the helpful discussions with X. Dai and X.C. Xie.
This work was financially supported by NSFC, Innovation project of
CAS and 973-project.

$^*$Email address: yupeng@aphy.iphy.ac.cn


\begin{thebibliography}{}
\bibitem{u} $n$ does not mean the true carrier
density in the present mean-field approach because we ignored $U$.
In the real materials the $T_c$ around $n=1$ could be depressed
heavily by $U$.
\end{thebibliography}
\end{document}